# Thermal expansion of boron subnitrides

Kirill A. Cherednichenko,[1] Lara Gigli [2] and Vladimir L. Solozhenko [1,*]

[1] *LSPM–CNRS, Université Paris Nord, 93430 Villetaneuse, France*

[2] *Elettra – Sincrotrone Trieste SCpA, 34149 Basovizza, Italy*

**Abstract**

The lattice parameters of two boron subnitrides, $B_{13}N_2$ and $B_{50}N_2$, have been measured as a function of temperature between 298 and 1273 K, and the corresponding thermal expansion coefficients have been determined. Thermal expansion of both boron subnitrides was found to be quasi-linear, and the volume thermal expansion coefficients of $B_{50}N_2$ ($15.7(2)\times10^{-6}$ $K^{-1}$) and $B_{13}N_2$ ($21.3(2)\times10^{-6}$ $K^{-1}$) are of the same order of magnitude as those of boron-rich compounds with structure related to α-rhombohedral boron. For both boron subnitrides no temperature-induced phase transitions have been observed in the temperature range under study.

## I. Introduction

Boron-rich solids are hard refractory compounds with superior thermal stability, excellent chemical resistance and outstanding mechanical properties [1]. Such fascinating combination of physical and chemical properties makes these materials promising for advanced superhard tooling solutions. High-speed machining leads to a high thermal load and considerable tool heating and may result in material failure. Thus, information on phase stability and thermal properties at high temperature is of crucial importance for producing new tool materials with increased productivity.

Rhombohedral $B_{13}N_2$ and tetragonal $B_{50}N_2$ boron subnitrides have been recently synthesized by crystallization from the B-BN melt at pressures of about 5 GPa [2-4]. Both subnitrides are refractory [4,5], low-compressible [6,7] and superhard [8,9] phases. In the present work thermal expansion of boron subnitrides has been studied in the 298-1273 K temperature range at ambient pressure by *in-situ* synchrotron X-ray powder diffraction.

---

* Corresponding author: vladimir.solozhenko@univ-paris13.fr

## II. Experimental

Powders of crystalline β-rhombohedral boron (99%, Alfa Aesar) and hexagonal graphite-like boron nitride (hBN) (99.8%, Johnson Matthey GmbH) were used as starting materials. Samples containing both boron subnitrides, $B_{13}N_2$ and $B_{50}N_2$, have been synthesized in a toroid-type apparatus with a specially designed high-temperature cell [9] at 5 GPa by quenching B–BN melt from 2630 K in accordance with high-pressure phase diagram of the B–BN system [10].

According to powder X-ray diffraction analysis (Equinox 1000 Inel diffractometer, Cu $K\alpha 1$ radiation) the recovered samples contain ~60 vol.% $B_{13}N_2$ and ~40 vol.% $B_{50}N_2$ with traces of hBN. Lattice parameters of $B_{13}N_2$ ($a = 5.4537(2)$ Å, $c = 12.2537(7)$ Å) and $B_{50}N_2$ ($a = 8.8181(2)$ Å, $c = 5.0427(10)$ Å) were calculated using Maud software [11] by least-squares fits to the indexed 2θ values; high purity silicon was used as an internal standard.

Thermal expansion of boron subnitrides has been studied at MCX beamline, Elettra Synchrotron (Trieste, Italy) in two independent experiments (Run 1 and Run 2). The powdered samples were loaded into quartz-glass capillaries under Ar atmosphere. Debye-Scherrer geometry with rotating capillary was used. The wavelength ($\lambda = 1.0352$ Å) was calibrated using Si as external standard, while temperature calibration was performed by measuring thermal expansion of platinum [12] under the same experimental conditions. X-ray diffraction patterns were collected in the 5-80 2θ-range using a translating image plate detector upon stepwise heating with 50-K steps from room temperature to 1273 K. Since the diffraction patterns collected in both runs are very similar, only 2D diagram from Run 1 is used to illustrate the experimental data (Fig. 1). The ramp time was 10 min with 5s of temperature stabilization, the acquisition time for each pattern was 120 s. The obtained X-ray diffraction data have been integrated into one-dimensional patterns using FIT2D software [13]. The lattice parameters were determined by Le Bail method using Powder Cell software [14]. At high temperatures the diffraction data were rather noisy, and only the most intense lines of $B_{13}N_2$ and $B_{50}N_2$ (marked in Fig. 1) have been used for Le Bail refinement.

## III. Results and Discussion

In spite of the fact that both boron subnitrides can be synthesized at the same *p-T* conditions, their structures are quite different (Fig. 2). As one can see, rhombohedral boron subnitride $B_{13}N_2$, has the structure related to α-rhombohedral boron and is isostructural to boron carbide, $B_{12}C_3$ i.e. $B_{12}$-icosahedra are placed on the corners of the rhombohedral unit cell, the interstitial N atoms are in 6*c* sites and boron atom is in 3*b* site [2]. Unlike $B_{13}N_2$, the structure of $B_{50}N_2$ is derived from α-tetragonal boron [15] and is built from $B_{12}$-icosahedra (each linked to six other $B_{12}$-units) and interstitial N–B–N chains [7]. According to the results of very recent compressibility study [7], $B_{50}N_2$ is significantly more compressible than other boron-rich compounds with structure related to α-rhombohedral boron.

Fig. 3 and Table I present the lattice parameters of both subnitrides at different temperatures. Since data from two independent experiments are in a reasonably good agreement, further we considered only the integral dataset (Run 1 + Run 2). The lattice parameters (*l*) of both subnitrides tend to increase in a monotonic way and *l*(T) curves (see Fig. 3) can be considered as quasi-linear ones. Hence, the corresponding linear expansion coefficients ($\alpha_l$) can be calculated using Eq. 1 [16]:

$$\alpha_l = \frac{l - l_{298\,K}}{l_{298\,K} \cdot (T - 298\,K)} \tag{1}$$

where $l_{298\,K}$ is the lattice parameter at room temperature. A significant anisotropy of unit cell thermal expansion has been observed for both subnitrides: $\alpha_c$-values are larger than the corresponding $\alpha_a$-values by 46% for $B_{13}N_2$ and by 34% for $B_{50}N_2$.

The average expansion coefficient ($\tilde{\alpha}$) can be calculated from the linear coefficients according to Eq. 2 [17]:

$$\tilde{\alpha} = \frac{1}{3}(2\alpha_a + \alpha_c) \tag{2}$$

The $\alpha_l$ and $\tilde{\alpha}$ values are presented in Table II. A significant anisotropy of the thermal expansion ($\alpha_c > \alpha_a$) observed for both boron subnitrides can be explained by the bonding features of interstitial nitrogen atoms. In $B_{50}N_2$ interstitial nitrogen atom is covalently bonded with four nearest icosahedra thus forming the distorted tetrahedron compressed in the *c*-direction. At high temperatures, the thermal vibrations due to their randomness result in the reduction of the tetrahedron distortion making the shorter *c*-parameter closer to the longer *a*-parameter similar to that in α-tetragonal boron [18]. In the case of $B_{13}N_2$, the bondings of nitrogen atom with boron atoms of the adjacent icosahedra lay in the *ab*-plane. Due to this, the elastic constant in the *c*-direction is lower than those in the *ab*-plane which is resulting in higher thermal expansion along *c*-axis, by analogy with $B_{13}C_2$ [19].

Unit cell volume changes of $B_{13}N_2$ and $B_{50}N_2$ in the 298-1273 K temperature range are shown in Fig. 4. One can see that thermal expansion of rhombohedral $B_{13}N_2$ is ~35% higher than that of tetragonal $B_{50}N_2$. The approximation of quasi-linear $V(T)/V_0$ dependencies has been done using Eq. 3 [20]:

$$V(T) = V_0\big(1 + \alpha_v(T - 298\,K)\big) \tag{3}$$

where $V_0$ is unit cell volume at ambient conditions; $\alpha_v$ is volume thermal expansion coefficient. The calculated volume thermal expansion coefficients of are listed in Table II.

The $\alpha_v$-values of boron subnitrides are of the same order of magnitude as volume thermal expansion coefficients of $B_{12}C_3$ ($17.2 \times 10^{-6}$ $K^{-1}$ [16], $21.2 \times 10^{-6}$ $K^{-1}$ [21], $17.6 \times 10^{-6}$ $K^{-1}$ [22]), $B_{12}O_2$ ($14.0 \times 10^{-6}$ $K^{-1}$ [20]), $B_{12}As_2$ ($15.0 \times 10^{-6}$ $K^{-1}$ [23]) and β-$B_{106}$ ($18.3 \times 10^{-6}$ $K^{-1}$ [24],

$19.2 \times 10^{-6}$ K$^{-1}$ [25]). Thus, among all boron-rich compounds with structures related to α-rhombohedral boron $B_{13}N_2$ is characterized by the highest volume thermal expansion coefficient. The explanation of this phenomenon should be a subject of separate theoretical study. On the other hand, one can see that the dispersion of experimental data on thermal expansion of isostructural boron-rich compounds is rather high, thus the additional precise high-temperature studies of various boron-rich compounds are highly demanded.

Particularly noteworthy is the fact that neither temperature-induced phase transitions nor decomposition are observed for both boron subnitrides in the temperature range under study.

## IV. Conclusions

Thermal expansion of two boron subnitrides, $B_{13}N_2$ and $B_{50}N_2$, has been studied *in situ* by synchrotron X-ray diffraction up to 1273 K at ambient pressure. The precise measurements of lattice parameters under heating allowed us to retrieve the thermal expansion coefficients of $B_{13}N_2$ and $B_{50}N_2$. The comparison of the experimentally found thermal expansion coefficients with corresponding literature data revealed that tetragonal $B_{50}N_2$ has one of the lowest $α_v$-values, while $α_v$-value of rhombohedral $B_{13}N_2$ is the highest among isostructural boron-rich compounds.

**Acknowledgements**

The authors thank Dr. Jasper Rikkert Plaisier and Giulio Zerauschek (Elettra) for assistance in synchrotron experiments, Dr. Thierry Chauveau (LSPM) for stimulating discussions on data analysis and an unknown reviewer for the valuable impact on the interpretation of thermal expansion anisotropy. X-ray diffraction experiments were carried out during beam time allocated for Proposal 20160086 at Elettra Sincrotrone Trieste. This work was financially supported by the European Union's Horizon 2020 Research and Innovation Programme under the Flintstone2020 project (grant agreement No. 689279).

Table I   Refined lattice parameters and unit cell volumes of boron subnitrides *versus* temperature at ambient pressure.

| Run 1 | | | | | | |
|---|---|---|---|---|---|---|
| T, K | $B_{13}N_2$ | | | $B_{50}N_2$ | | |
|  | $a$, Å | $c$, Å | V, Å$^3$ | $a$, Å | $c$, Å | V, Å$^3$ |
| 298 | 5.4537(3) | 12.2537(7) | 315.6(1) | 8.8181(2) | 5.0427(10) | 392.1(1) |
| 323 | 5.4552(2) | 12.2605(6) | 315.9(1) | 8.8177(1) | 5.0447(6) | 392.2(1) |
| 373 | 5.4563(3) | 12.2665(6) | 316.3(1) | 8.8185(2) | 5.0447(5) | 392.3(1) |
| 423 | 5.4567(2) | 12.2695(7) | 316.4(0) | 8.8205(2) | 5.0455(4) | 392.6(1) |
| 473 | 5.4591(3) | 12.2843(5) | 317.1(1) | 8.8242(1) | 5.0482(4) | 393.1(1) |
| 523 | 5.4616(4) | 12.2845(6) | 317.4(1) | 8.8272(2) | 5.0499(4) | 393.5(1) |
| 573 | 5.4619(4) | 12.2883(7) | 317.5(1) | 8.8295(3) | 5.0503(4) | 393.7(1) |
| 623 | 5.4625(5) | 12.2913(7) | 317.6(1) | 8.8321(3) | 5.0519(5) | 394.1(1) |
| 673 | 5.4644(4) | 12.2955(8) | 317.9(1) | 8.8341(3) | 5.0530(5) | 394.3(1) |
| 723 | 5.4660(5) | 12.2994(7) | 318.2(1) | 8.8370(2) | 5.0551(5) | 394.8(1) |
| 773 | 5.4683(5) | 12.3076(8) | 318.7(1) | 8.8390(3) | 5.0570(6) | 395.1(1) |
| 823 | 5.4696(4) | 12.3093(9) | 318.9(1) | 8.8443(4) | 5.0620(6) | 395.9(1) |
| 873 | 5.4699(3) | 12.3153(9) | 319.1(1) | 8.8450(4) | 5.0625(7) | 396.1(1) |
| 923 | 5.4725(5) | 12.3205(10) | 319.5(1) | 8.8465(5) | 5.0647(7) | 396.4(1) |
| 973 | 5.4747(6) | 12.3259(11) | 319.9(1) | 8.8477(4) | 5.0658(7) | 396.6(2) |
| 1023 | 5.4761(6) | 12.3249(12) | 320.1(1) | 8.8495(5) | 5.0666(8) | 396.8(2) |
| 1073 | 5.4777(7) | 12.3339(10) | 320.5(1) | 8.8523(6) | 5.0684(8) | 397.2(2) |
| 1123 | 5.4807(6) | 12.3456(13) | 321.2(1) | 8.8523(6) | 5.0692(9) | 397.2(2) |
| 1173 | 5.4807(7) | 12.3375(11) | 320.9(1) | 8.8542(7) | 5.0700(10) | 397.5(2) |
| 1223 | 5.4823(8) | 12.3511(12) | 321.5(1) | 8.8568(8) | 5.0709(13) | 397.8(2) |
| Run 2 | | | | | | |
| T, K | $B_{13}N_2$ | | | $B_{50}N_2$ | | |
|  | $a$, Å | $c$, Å | V, Å$^3$ | $a$, Å | $c$, Å | V, Å$^3$ |
| 298 | 5.4537(3) | 12.2537(7) | 315.6(1) | 8.8181(2) | 5.0427(10) | 392.1(1) |
| 323 | 5.4538(3) | 12.2572(5) | 315.7(1) | 8.8190(4) | 5.0432(9) | 392.2(1) |
| 373 | 5.4557(2) | 12.2679(6) | 316.2(0) | 8.8199(3) | 5.0444(8) | 392.4(1) |
| 423 | 5.4560(3) | 12.2720(6) | 316.4(1) | 8.8230(2) | 5.0465(7) | 392.8(1) |
| 473 | 5.4590(2) | 12.2800(7) | 316.9(0) | 8.8238(4) | 5.0468(6) | 392.9(1) |
| 523 | 5.4604(3) | 12.2842(6) | 317.2(1) | 8.8277(6) | 5.0496(5) | 393.5(1) |
| 573 | 5.4609(4) | 12.2840(7) | 317.3(1) | 8.8310(5) | 5.0498(7) | 393.8(1) |
| 623 | 5.4645(4) | 12.2960(8) | 317.9(1) | 8.8320(4) | 5.0512(9) | 394.0(1) |
| 673 | 5.4664(5) | 12.2998(7) | 318.3(1) | 8.8325(5) | 5.0549(8) | 394.4(1) |
| 723 | 5.4681(4) | 12.3038(7) | 318.6(1) | 8.8370(6) | 5.0551(7) | 394.8(1) |
| 773 | 5.4720(5) | 12.3157(7) | 319.4(1) | 8.8384(6) | 5.0580(10) | 395.1(1) |
| 823 | 5.4730(5) | 12.3190(8) | 319.6(1) | 8.8400(6) | 5.0584(9) | 395.3(1) |
| 873 | 5.4745(4) | 12.3197(7) | 319.8(1) | 8.8390(5) | 5.0589(12) | 395.2(1) |
| 923 | 5.4769(3) | 12.3255(7) | 320.2(1) | 8.8430(7) | 5.0616(10) | 395.8(1) |
| 973 | 5.4778(5) | 12.3290(8) | 320.4(1) | 8.8470(6) | 5.0663(9) | 396.5(1) |
| 1023 | 5.4806(6) | 12.3337(9) | 320.8(1) | 8.8450(8) | 5.0663(13) | 396.4(1) |
| 1073 | 5.4825(6) | 12.3390(9) | 321.2(1) | 8.8490(9) | 5.0684(11) | 396.9(1) |
| 1123 | 5.4846(7) | 12.3430(10) | 321.5(1) | 8.8500(8) | 5.0682(12) | 396.9(1) |
| 1173 | 5.4846(6) | 12.3447(11) | 321.6(1) | 8.8531(9) | 5.0719(10) | 397.5(1) |
| 1223 | 5.4865(7) | 12.3526(10) | 322.0(1) | 8.8538(11) | 5.0717(12) | 397.6(1) |
| 1273 | 5.4869(8) | 12.3650(13) | 322.4(1) | 8.8559(10) | 5.0727(10) | 397.8(2) |

Table II. Linear ($\alpha_l$), average ($\tilde{\alpha}$) and volume ($\alpha_v$) thermal expansion coefficients of boron subnitrides.

|  | $\alpha_a \times 10^6$, K$^{-1}$ | $\alpha_c \times 10^6$, K$^{-1}$ | $\tilde{\alpha} \times 10^6$, K$^{-1}$ | $\alpha_v \times 10^6$, K$^{-1}$ |
|---|---|---|---|---|
| B$_{13}$N$_2$ | 6.1(1) | 8.9(1) | 7.0(1) | 21.3(2) |
| B$_{50}$N$_2$ | 4.7(1) | 6.3(1) | 5.2(1) | 15.7(2) |

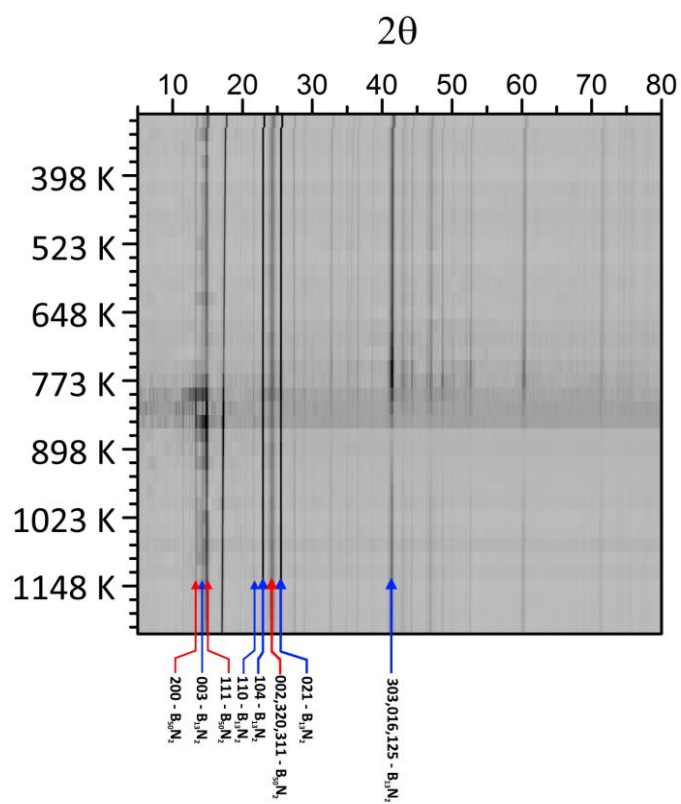

Fig. 1. 2D diagram of X-ray diffraction patterns of boron subnitrides *versus* temperature in Run 1. The most intense lines of $B_{13}N_2$ and $B_{50}N_2$ which have been used for Le Bail refinement are marked by blue and red arrows, respectively.

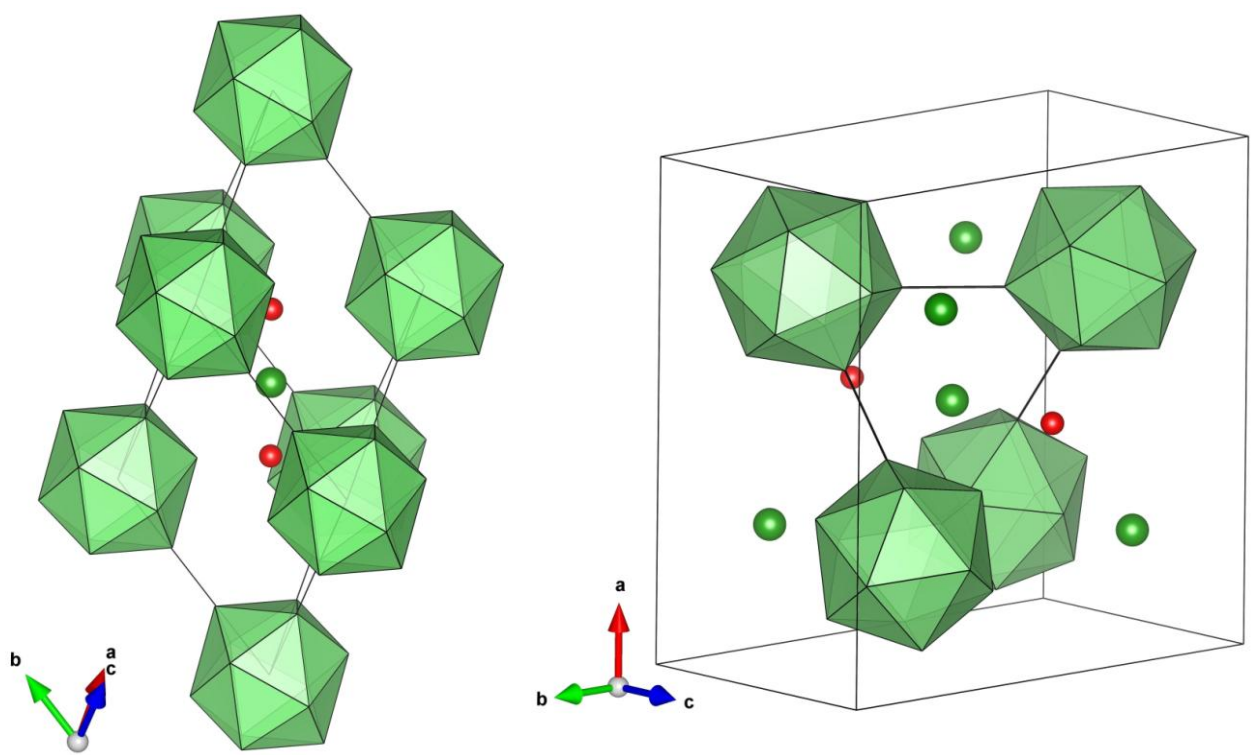

Fig. 2. Crystal structures of rhombohedral $B_{13}N_2$ [2] (*left*) and tetragonal $B_{50}N_2$ [7] (*right*). Nitrogen and boron atoms are presented in red and green, respectively.

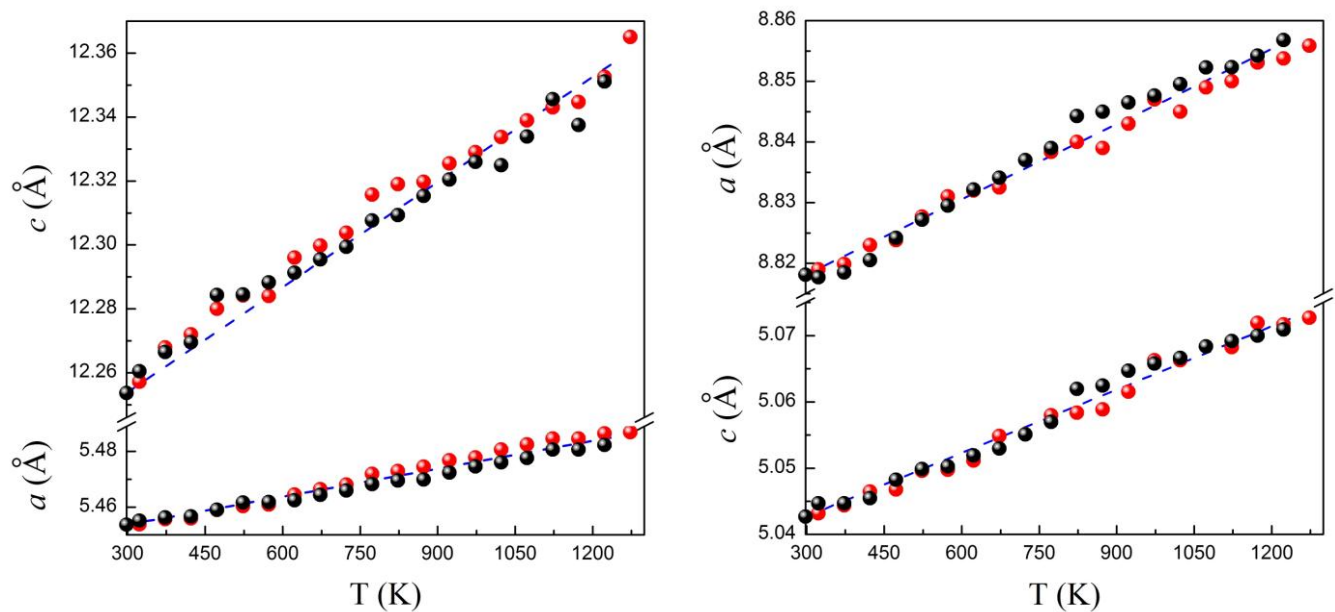

Fig. 3. Lattice parameters of $B_{13}N_2$ (*left*) and $B_{50}N_2$ (*right*) as a function of temperature. The dashed lines represent the corresponding linear fits to the experimental data obtained in Run 1 and Run 2 (black and red circles, respectively).

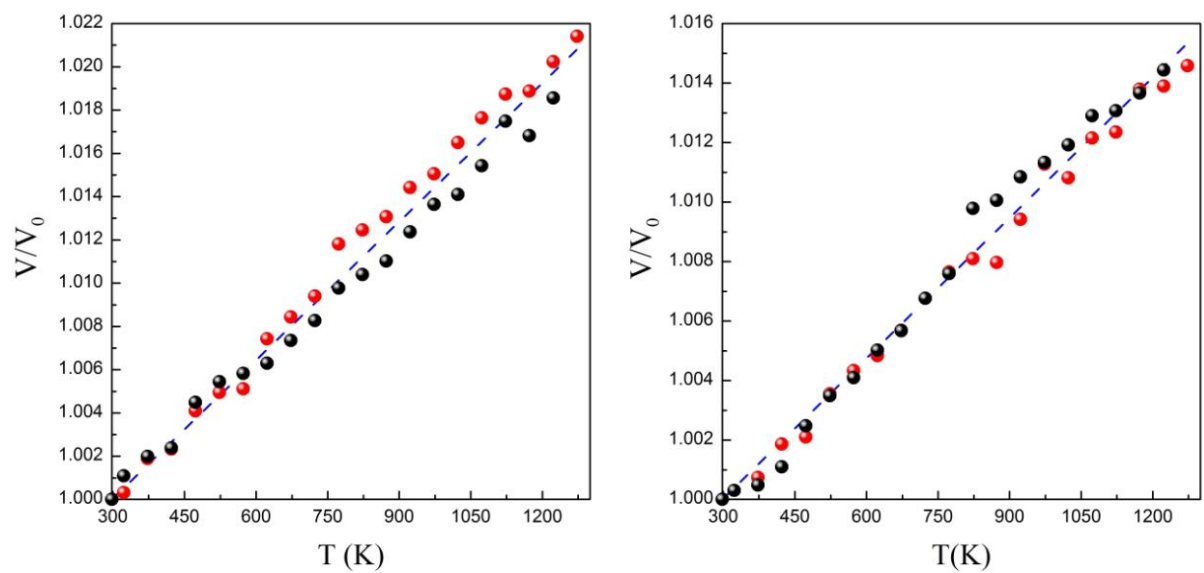

Fig. 4. Normalized unit cell volumes of $B_{13}N_2$ (*left*) and $B_{50}N_2$ (*right*) as a function of temperature. The dashed lines represent the linear fits to the experimental data obtained in Run 1 and Run 2 (black and red circles, respectively).